\documentclass[12pt]{article}

\usepackage{graphicx}
\usepackage{dcolumn}
\usepackage{bm}
\usepackage{amsmath}
\usepackage{epstopdf}
\usepackage{amsfonts}
\usepackage{amssymb}
\usepackage{epsfig}
\usepackage{tabularx}
\usepackage{color}

\usepackage[colorlinks=true,citecolor=blue,urlcolor=magenta,breaklinks]{hyperref}

\newcommand{\be}{\begin{equation}}
\newcommand{\en}{\end{equation}}
\newcommand{\bea}{\begin{eqnarray}}
\newcommand{\ena}{\end{eqnarray}}

\begin{document}

\begin{titlepage}

%\rightline{hep-th/yymmnnn}

%\vskip 2cm

\centerline{\large \bf {Quasinormal modes of charged black holes in}}

\centerline{\large \bf {higher-dimensional Einstein-power-Maxwell theory}}

\vskip 1.5cm

\centerline{Grigoris Panotopoulos}

\vskip 1cm

\centerline{Centro de Astrof{\'i}sica e Gravita\c c\~ao-CENTRA, Departamento de F{\'i}sica,} 

\centerline{Instituto Superior T{\'e}cnico-IST, Universidade de Lisboa-UL,}

\centerline{Avenida Rovisco Pais 1, 1049-001, Lisboa, Portugal}

\centerline{email:
\href{mailto:grigorios.panotopoulos@tecnico.ulisboa.pt}
{\nolinkurl{grigorios.panotopoulos@tecnico.ulisboa.pt}}
}

\vskip 1.5cm

\begin{abstract}
We compute the quasinormal frequencies for scalar perturbations of charged black holes in five-dimensional Einstein-power-Maxwell theory. The impact on the spectrum of the electric charge of the black holes, of the angular degree, of the overtone number, and of the mass of the test scalar field is investigated in detail. The quasinormal spectra in the eikonal limit are computed as well for several different space-time dimensionalities.
\end{abstract}

\end{titlepage}

%%%%%%%%%%%%%%%%%%%%%%%%
\section{Introduction}
%%%%%%%%%%%%%%%%%%%%%%%%

In all metric theories of gravity black holes (BHs) generically are predicted to exist.
Despite their simplicity are without a doubt fascinating objects both for classical and quantum gravity. Although they are characterized by a handful of parameters, such as mass, spin, and charges (even if one includes BHs with a scalar hair \cite{carlos}), black holes link together several research areas, from gravitation and statistical physics to quantum mechanics and Astrophysics. By now their existence has been established in a two-fold way, namely first by the numerous direct detections of gravitational waves from black hole mergers \cite{ligo1,ligo2,ligo3,ligo4,ligo5}, and then thanks to the first image of the supermassive black hole \cite{L1,L2,L3,L4,L5,L6} located at the centre of the giant elliptical galaxy Messier 87 by the Event Horizon Telescope project \cite{project} precisely one year ago.

Isolated black holes are in fact ideal objects. It turns out that realistic black holes in Nature are never isolated. Instead, they are constantly affected by their environment. Upon external perturbations black holes respond by emitting gravitational waves. Quasinormal modes (QNMs) are characteristic frequencies that depend on the details of the geometry and on the type of the perturbation (scalar, Dirac, vector or tensor), but not on the initial conditions.
The information on how BHs relax after the perturbation has been applied is encoded into the
non-vanishing imaginary part of the QN frequencies. Black hole perturbation theory \cite{wheeler,zerilli1,zerilli2,zerilli3,moncrief,teukolsky} and QNMs become relevant during the ring down phase of a black hole merger, the stage where a single distorted object is formed, and where the geometry of space time undergoes dumped oscillations due to the emission of gravitational waves. Taking advantage of gravitational wave Astronomy we have now a powerful tool at our disposal that allows us to test gravitation under extreme conditions, such as BH mimickers, modifications of gravity, the Kerr paradigm of GR etc. For excellent reviews on the topic see \cite{review1,review2,review3}, and also the Chandrasekhar's monograph \cite{monograph}, which is the standard textbook on the mathematical aspects of black holes. 

Another issue regarding BH QNMs is the so-called "Bohr-like approach to black hole quantum physics". As the name suggests, that approach is somewhat similar to the historical semi-classical model of the structure of a hydrogen atom introduced by Bohr in 1913 \cite{Bohr1,Bohr2}. In that approach, which has been published in \cite{corda1,corda2} and then reviewed in \cite{corda3}, in a certain sense, the QNMs "triggered" by the emissions of Hawking quanta and by the absorption of external particles represent the "electron" which jumps from one level to another, and the absolute values of the quasi-normal frequencies represent the energy "shells". Finally, it is worth mentioning here that the aforementioned approach proposes also an interesting solution to the BH information paradox \cite{corda4}.

A special attention has been devoted to non-linear electrodynamics (NLE) for several decades now, which has a long history and it has been studied over the years in several different contexts. To begin with, it is known that classical electrodynamics is based on a system of linear equations. However, when quantum effects are taken into account, the effective equations become non-linear. The first models go many decades back, when Euler and Heisenberg calculated QED corrections \cite{Euler}, while Born and Infeld managed to obtain a finite self-energy of point-like charges \cite{BI} in the 30's. Moreover, assuming appropriate non-linear electromagnetic sources, which in the weak field limit are reduced to the usual Maxwell's linear theory, one can generate a new class of Bardeen-like \cite{Bardeen,borde} BH solutions  \cite{beato1,beato2,beato3,bronnikov,dymnikova,hayward,vagenas1,vagenas2,Pallikaris}. Those solutions on the one hand do have a horizon, and on the other hand their curvature invariants, such as the Ricci scalar $R$, are regular everywhere, contrary to the standard Reissner-Nordstr{\"o}m solution \cite{RN}. 

Finally, a simple model that generalizes Maxwell's linear theory in a straightforward manner leads to the so called Einstein-power-Maxwell (EpM) theory \cite{EpM1,EpM2,EpM3,EpM4,EpM5,EpM6,EpM7,EpM8,EpM9,EpM10,EpM11}, described by a Lagrangian density of the form $\mathcal{L}(F) \sim F^k$, where $F$ is the Maxwell invariant, and $k$ is an arbitrary rational number. Although our observable Universe is clearly four-dimensional, the question "How many dimensions are there?" is one of the fundamental questions that modern High Energy Physics tries to answer. Kaluza-Klein theories \cite{kaluza,klein}, Supergravity \cite{nilles} and Superstring/M-Theory \cite{ST1,ST2} have pushed forward the idea that extra spatial dimensions may exist. The advantage of the EpM theory is that it preserves the nice conformal properties of the four-dimensional Maxwell's theory in any number of space-time dimensionality $D$, provided that the power $k$ is chosen to be $k=D/4$, as it is easy to verify that for this particular value the electromagnetic stress-energy tensor becomes traceless.

What is more, over the years the computation of QNMs of higher-dimensional BHs has attracted a lot of attention for several reasons, such as i) the study of features of higher-dimensional GR \cite{extra1,extra2}, ii) the analysis of the physical implications of the brane-world scenario \cite{kanti}, and iii) the understanding of thermodynamic properties of BHs in Loop Quantum Gravity \cite{extra3,extra4}, to mention just a few. Given the interest in Gravitational Wave Astronomy and in QNMs of black holes, it would be interesting to see what kind of QN spectra are expected from higher-dimensional black holes in EpM theory. In this work we propose to compute the QN spectrum for scalar perturbations of charged BHs in EpM theory in higher dimensions, extending two previous similar works of ours \cite{EpM8,EpM11} in $D > 4$.

Our work in the present article is organized as follows: In the next section we briefly review charged BH solutions in EpM theory, and we also very briefly discuss the wave equation with the corresponding effective potential barrier for the scalar perturbations. In the third section we compute the quasinormal frequencies adopting the WKB approximation of 6th order, and we discuss our results. There we also compute the QNMs in the eikonal regime for several different values of space-time dimensionality. Finally, in section four we summarize our work with some concluding remarks. We adopt the mostly positive metric signature $(-,+,+,+)$, and we work in geometrical units where the universal constants are set to unity, $c=1=G_5$.

%%%%%%%%%%%%%%%%%%%%%%%%%%%%%%%%%%%%%%%%%%%%%%%%%%%%%%%%%%%%%%%%%%%%%%%%%%%
\section{The black hole gravitational background and scalar perturbations}
%%%%%%%%%%%%%%%%%%%%%%%%%%%%%%%%%%%%%%%%%%%%%%%%%%%%%%%%%%%%%%%%%%%%%%%%%%%

\subsection{Charged black hole solutions in EpM theory}

We are interested in a D-dimensional theory described by the action
\begin{equation}
S[g_{\mu \nu}, A_\mu] = \int \mathrm{d} ^Dx \sqrt{-g} \left[ \frac{1}{2 \kappa} R - \alpha (F_{\mu \nu} F^{\mu \nu})^k \right],
\end{equation}
where $R$ is the Ricci scalar, $g$ the determinant of the metric tensor $g_{\mu \nu}$, $A_\mu$ is the Maxwell potential, $k$ is an arbitrary rational number, $\kappa=8 \pi$ and $F \equiv F_{\mu \nu} F^{\mu \nu}$ the Maxwell invariant with $F_{\mu \nu}$ being the electromagnetic field strength defined by 
\begin{equation}
F_{\mu \nu} \equiv \partial_\mu A_\nu - \partial_\nu A_\mu
\end{equation}
where the indices run from 0 to $D-1$. 

Varying the action with respect to the metric tensor one obtains Einstein's field equations sourced by the electromagnetic energy-momentum tensor \cite{EpM1}
\begin{equation}
G_{\mu \nu}  =  4 \kappa \alpha \left [k F_{\mu \rho} F_\nu ^\rho F^{k-1} - \frac{1}{4} g_{\mu \nu} F^k \right ],
\end{equation}
with $G_{\mu \nu}$ being the Einstein tensor. Next, varying the action with respect to the Maxwell potential $A_\mu$ one obtains the generalized Maxwell equations \cite{EpM1}
\begin{equation}
\partial_\mu (\sqrt{-g} F^{\mu \nu} F^{k-1}) = 0.
\end{equation}
We seek static, spherically symmetric solutions making as usual for the metric tensor the Ans{\"a}tz
\begin{equation}
ds^2 = -f(r) dt^2 + f(r)^{-1} dr^2 + r^2 d \Omega_{D-2}^2,
\end{equation}
with $r$ being the radial coordinate, and with $d \Omega_{D-2}^2$ being the line element of the unit (D-2)-dimensional sphere. In the case $D=5$, for instance, it is given by \cite{EGB1,EGB2}
\begin{equation}
\mathrm{d} \Omega_3^2 = \mathrm{d} \theta^2 + \sin^2 \theta \mathrm{d} \phi^2 + \sin^2 \theta \sin^2 \phi \mathrm{d}\psi^2
\end{equation}
The electric field $E(r)$ is found to be \cite{EpM1}
\begin{equation}
E(r)=F_{rt} = \frac{C}{r^\beta},
\end{equation}
where $C$ is a constant of integration, and the power $\beta$ is given by \cite{EpM1}
\begin{equation}
\beta = \frac{2 (D k-4 k+1)}{2k-1},
\end{equation}
while the metric function $f(r)$ is computed to be \cite{EpM1}
\begin{equation}
f(r) = 1-\frac{\mu}{r^{D-3}}+\frac{q}{r^\beta},
\end{equation}
where $\mu, q$ are two constants related to the mass $M$ and the electric charge $Q$ of the BH, respectively. The constant $q$ and the constant of integration $C$ in five dimensions are related via \cite{EpM1}
\begin{equation}
q = - 2 \kappa \alpha (-2 C^2)^{k}  \: \frac{(2k-1)^2}{(D-2) (D-2 k-1)}.
\end{equation}
Furthermore, we shall consider the case in which the power $\beta > 2$. In order to have real roots for the metric function $f(r)$, the constants $\mu,q$ must satisfy the conditions $\mu > 0$ and \cite{EpM1}
\begin{equation}
0 < q < q_{max}
\end{equation}
where the upper bound corresponds to extremal BHs, and it is given by
\begin{equation}
q_{max} = (D-3) \left( \frac{\mu}{\beta} \right)^{\beta/(D-3)} (\beta+3-D)^{\frac{\beta+3-D}{D-3}}
\end{equation}
Clearly, setting $k=1$ we recover the well-known case of the higher-dimensional version of the Reissner-Nordstr{\"o}m BH \cite{RN} of Maxwell's linear electrodynamics. Furthermore, setting $q=0$ the solution for the higher-dimensional version of the Schwarzschild geometry \cite{Tangherlini} is recovered.

%%%%%%%%%%%%%%%%%%%%%%%%%%%%%%%%%%%%%%%%%%%%%%%%%%%%%%
\subsection{Wave equation for scalar perturbations}
%%%%%%%%%%%%%%%%%%%%%%%%%%%%%%%%%%%%%%%%%%%%%%%%%%%%%

Here we very briefly review the necessary ingredients to deal with the computation of the QN frequencies for scalar perturbations. To that end, let us consider the propagation of a test scalar field, $\Phi$, which is assumed to be real, massive, electrically neutral, and canonically coupled to gravity, in a fixed gravitational background. Its wave equation is given by the usual Klein-Gordon equation, see e.g. \cite{crispino,Pappas1,Pappas2,KGEquation}
\begin{equation}
\frac{1}{\sqrt{-g}} \partial_\mu (\sqrt{-g} g^{\mu \nu} \partial_\nu) \Phi = m^2 \Phi
\end{equation}
with $m$ being the mass of the test scalar field.

In order to solve the Klein-Gordon equation, we apply as usual the separation of variables making the following Ans{\"a}tz:
\begin{equation}\label{separable}
\Phi(t,r,\theta,\phi,\psi,...) = e^{-i \omega t} \: \frac{y(r)}{r^{(D-2)/2}} \: \tilde{Y}_l (\Omega)
\end{equation}
where $\omega$ is the frequency to be determined, while $\tilde{Y}_l(\Omega)$ is the higher-dimensional generalization of the usual spherical harmonics, and they depend on the angular coordinates \cite{book}. 
Making the previous Ans{\"a}tz it is straightforward to obtain for the radial part a Schr{\"o}dinger-like equation
\begin{equation}
\frac{\mathrm{d}^2 y}{\mathrm{d}x^2} + [ \omega^2 - V(x) ] \: y = 0
\end{equation}
with $x$ being the tortoise coordinate, i.e.,
\begin{equation}
x  \equiv  \int \frac{\mathrm{d}r}{f(r)}
\end{equation}
Finally, the effective potential barrier for scalar perturbations in any number of dimensions $D$ is given by \cite{ref}
\begin{equation}
V(r) = f(r) \: \left(m^2 + \frac{l (l+D-3)}{r^2} + \frac{D-2}{2} \: \frac{f'(r)}{r} + \frac{(D-2) (D-4)}{4} \: \frac{f(r)}{r^2} \right)
\end{equation}
where the prime denotes differentiation with respect to $r$, and $l \geq 0$ is the angular degree.

\begin{figure}[ht!]
\centering
\includegraphics[scale=0.9]{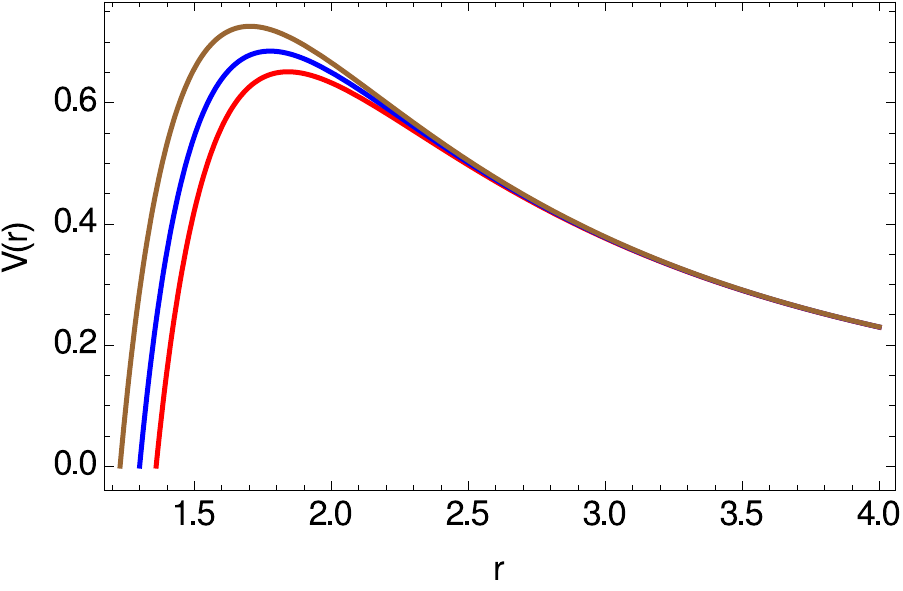} \
\includegraphics[scale=0.9]{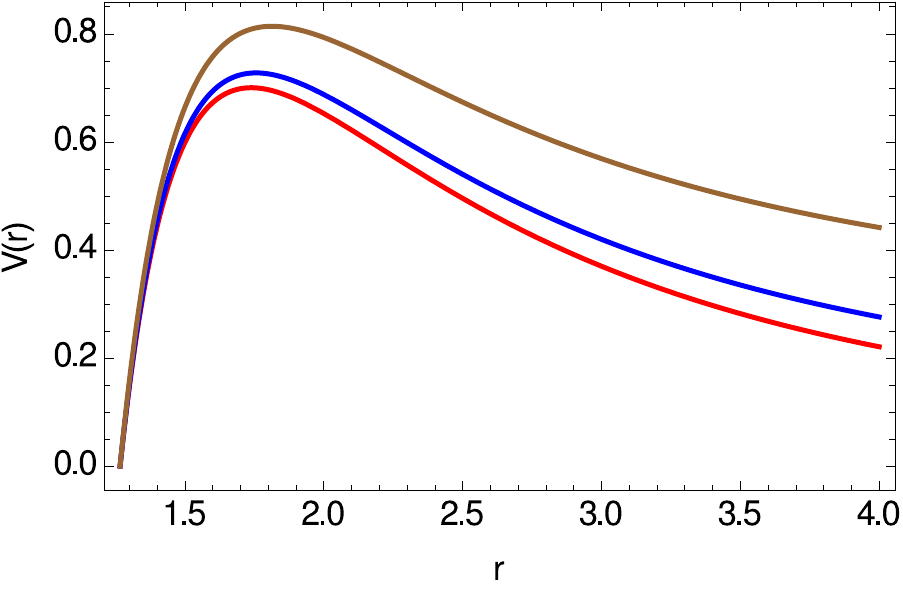}
\caption{
Effective potential barrier for scalar perturbations vs radial coordinate for $D=5$, $k=5/4$, $\mu=2$ and $l=1$. {\bf Top panel:} From bottom to top $q=0.2$ (red), $q=0.4$ (blue) and $q=0.6$ (brown).
{\bf Bottom panel:} From bottom to top $m=0$ (red), $m=0.25$ (blue) and $m=0.5$ (brown).
}
\label{fig:1} 	
\end{figure}

The effective potential barrier for $D=5$ and $k=5/4$ is shown in Fig.~\ref{fig:1} for three different values of $q$ (top panel) and three different values of $m$ (bottom panel) for fixed $\mu=2,l=1$.

%%%%%%%%%%%%%%%%%%%%%%%%%%%%%%%%%%
\section{Quasinormal frequencies}
%%%%%%%%%%%%%%%%%%%%%%%%%%%%%%%%%%

In the discussion to follow, we shall fix the mass parameter of the BHs $\mu=2$, and for concreteness we shall consider the five-dimensional case
\begin{equation}
D=5, \quad k=5/4, \quad \beta=3
\end{equation}
since as already mentioned in the introduction this is precisely the value of the power $k$ for which the electromagnetic stress-energy tensor becomes traceless. Then the allowed range for $q$ becomes, $0 < q < q_{max} \simeq 1.09$. In the next subsection we shall compare the QN spectra in the eikonal limit for several different space-time dimensionalities.

Exact analytical calculations for QN spectra of black holes may be performed in some special cases, e.g. either when the effective potential barrier takes the form of the P{\"o}schl-Teller potential \cite{potential,ferrari,cardoso2,lemos,molina,panotop1}, or when the differential equation for the radial part of the wave function may be recast into the Gauss' hypergeometric function \cite{exact1,exact2,exact3,exact4,exact5,exact6}. More generically, however, due to the complexity of the problem, some numerical method is employed. Over the years several different methods to compute the QNMs of black holes have been developed, such as the Frobenius method, generalization of the Frobenius series, fit and interpolation approach, method of continued fraction etc. For more details the interested reader may consult e.g. \cite{review3}. In particular, semi-analytical methods based on the Wentzel-Kramers-Brillouin (WKB) approximation (familiar from non-relativistic quantum mechanics) \cite{wkb1,wkb2,wkb3} are among the most popular ones, and they have been extensively applied to several cases. For a partial list see for instance \cite{paper1,paper2,paper3,paper4,paper5,paper6}, and for more recent works \cite{paper7,paper8,paper9,paper10,Rincon:2018sgd}, and references therein.

Applying the WKB method, the QN spectra may be computed making use of the following expression
\begin{equation}
\omega_n^2 = V_0+(-2V_0'')^{1/2} \Lambda(n) - i \nu (-2V_0'')^{1/2} [1+\Omega(n)]
\end{equation}
where $n=0,1,2...$ is the overtone number, $\nu=n+1/2$, $V_0$ is the maximum of the effective potential barrier, $V_0''$ is the second derivative of the potential evaluated at the maximum, while $\Lambda(n), \Omega(n)$ are complicated expressions of $\nu$ and higher derivatives of the potential evaluated at the maximum, and they can be found e.g. in \cite{paper2,paper7}. 
Here we have used the Wolfram Mathematica \cite{wolfram} code with WKB at any order from one to six presented in \cite{code}. For a given angular degree $l$ we have considered values $n < l$ only, since this is the case for which the best results are obtained, see e.g. Tables II, III, IV and V of \cite{Opala}.
For higher order WKB corrections, and recipes for simple, quick, efficient and accurate computations see \cite{Opala,Konoplya:2019hlu,RefExtra2}.

\begin{figure}[ht!]
\centering
\includegraphics[scale=0.9]{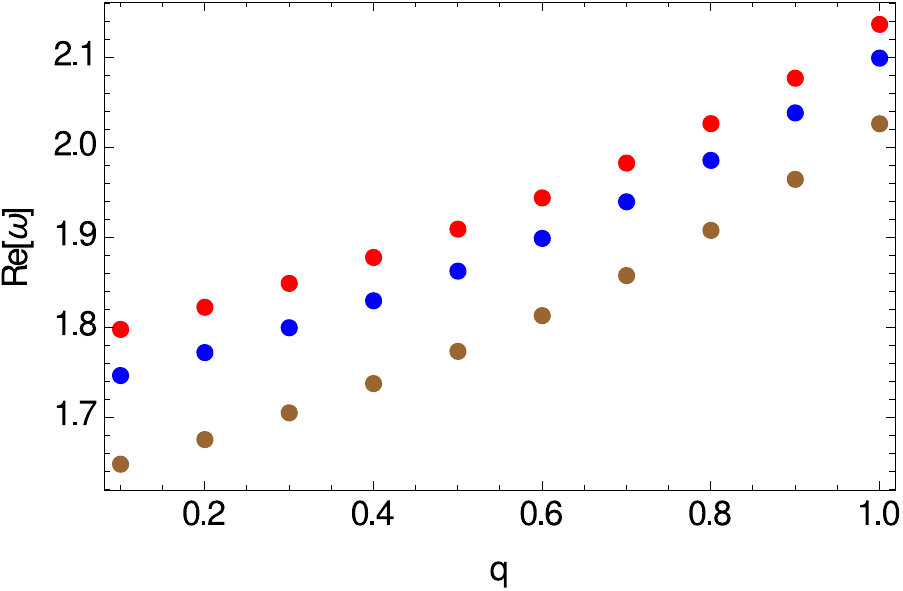} \
\includegraphics[scale=0.9]{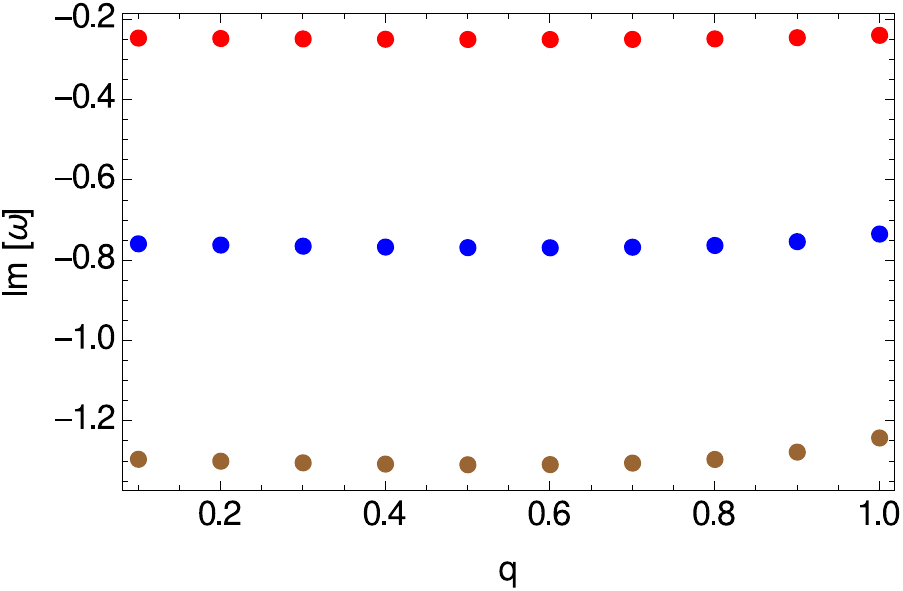}
\caption{
Real part (top panel) and imaginary part (bottom panel) of the modes vs 
electric charge for $D=5, k=5/4, \mu=2, m=0.001, l=4$. From top to down $n=0,1,2$.
}
\label{fig:2} 	
\end{figure}

\begin{figure}[ht!]
\centering
\includegraphics[scale=0.9]{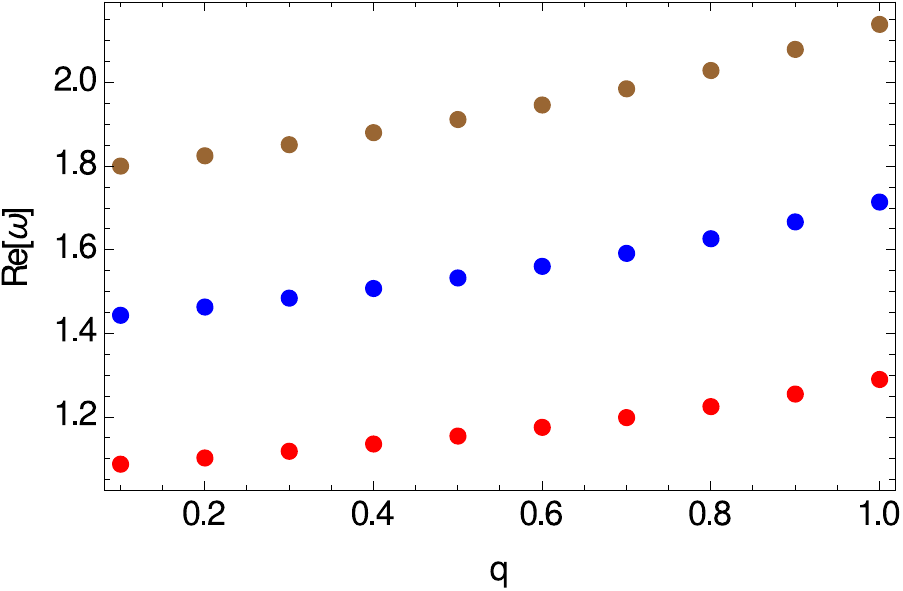} \
\includegraphics[scale=0.9]{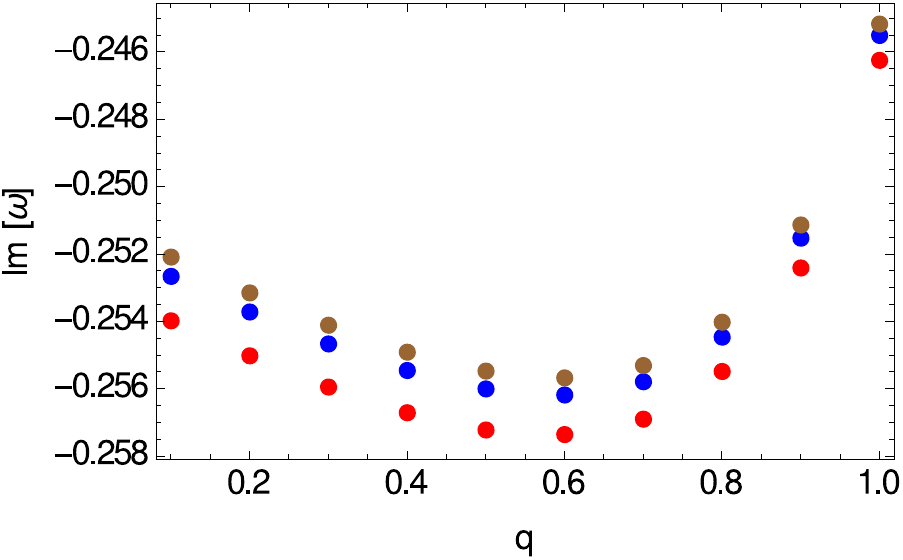}
\caption{
Real part (top panel) and imaginary part (bottom panel) of the modes vs 
electric charge for $D=5, k=5/4, \mu=2, m=0.001, n=0$. From bottom to top $l=2,3,4$.
}
\label{fig:3} 	
\end{figure}

\begin{figure}[ht!]
\centering
\includegraphics[scale=0.9]{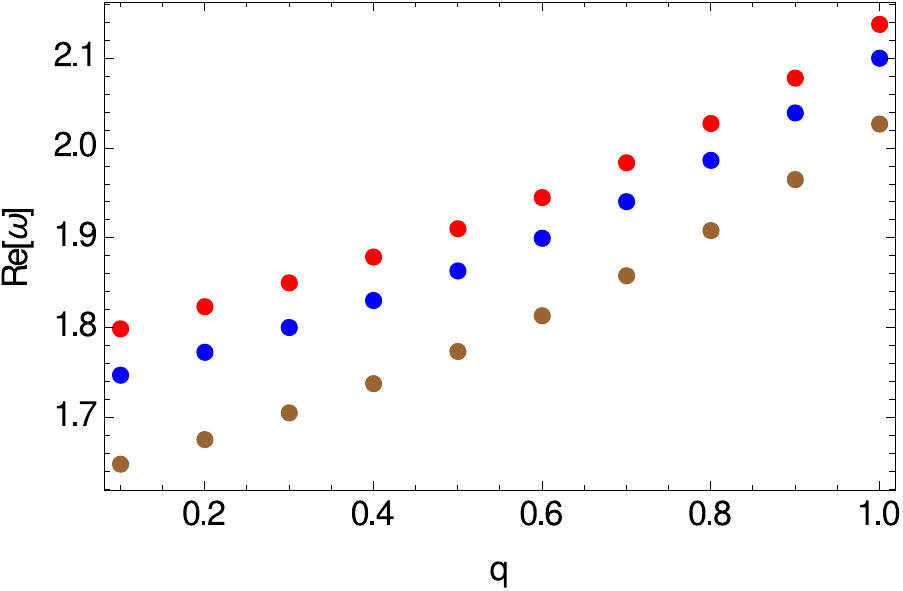} \
\includegraphics[scale=0.9]{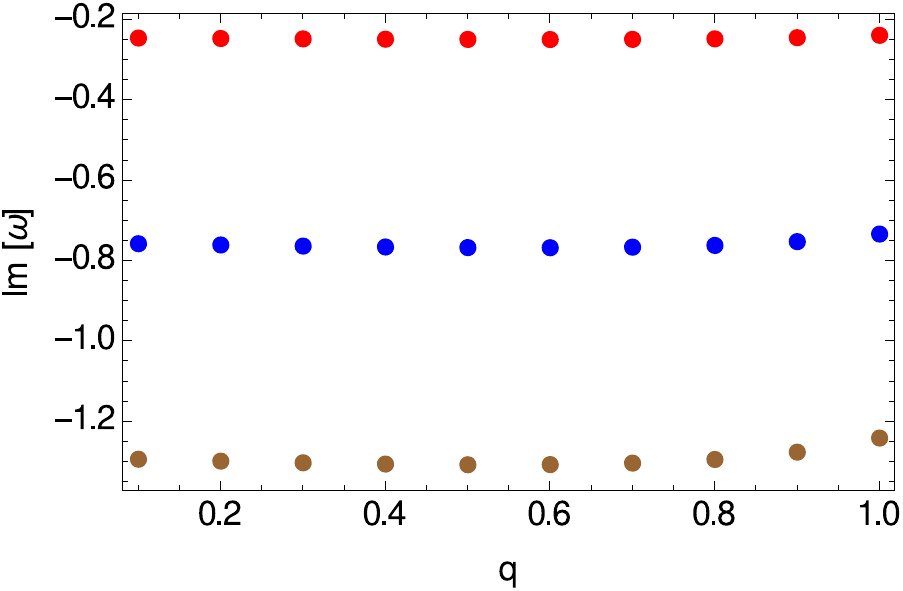}
\caption{
Same as Fig.~\ref{fig:2}, but for $m=0.1$.
}
\label{fig:4} 	
\end{figure}

\begin{figure}[ht!]
\centering
\includegraphics[scale=0.9]{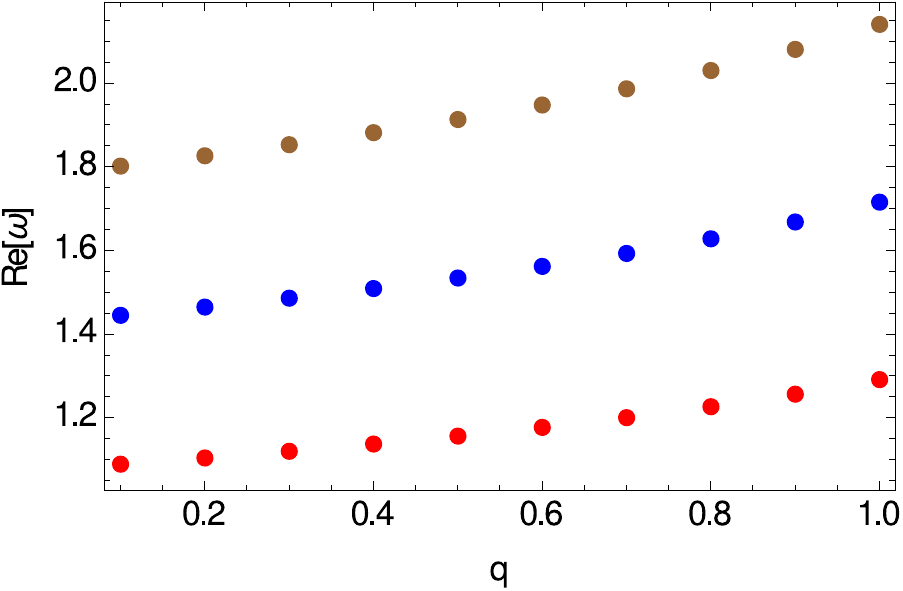} \
\includegraphics[scale=0.96]{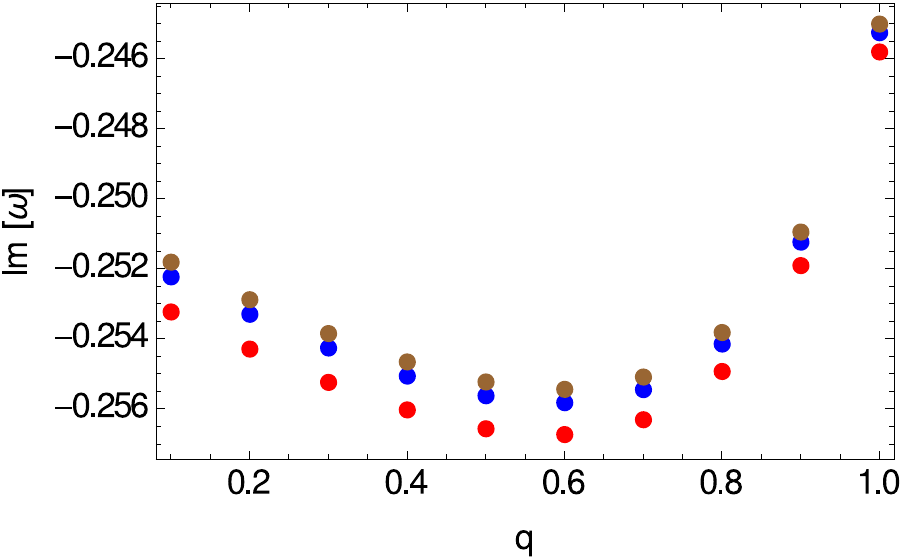}
\caption{
Same as Fig.~\ref{fig:3}, but for $m=0.1$.
}
\label{fig:5} 	
\end{figure}

Our main numerical results are summarized in Fig.~\ref{fig:2}, \ref{fig:3}, \ref{fig:4} and \ref{fig:5}. In particular, in Fig.~\ref{fig:2} we show the real and the imaginary part of the frequencies vs the electric charge of the BHs, $q$, for $\mu=2,m=0.001,l=4$ and three different values of the overtone number, $n=0,1,2$. In Fig.~\ref{fig:3} we show the real and the imaginary part of the modes vs the electric charge of the BHs for $\mu=2,m=0.001,n=0$ and three different values of the angular degree, $l=2,3,4$. In the other two plots we have considered a heavier test scalar field with a mass $m=0.1$. Although it is not visible in the plots shown here, there is a small difference in the values between the $m=0.001$ and the $m=0.1$ case. This is demonstrated in Table~\ref{table:FirstSet} for the case $l=4, n=0$. Overall we can say that the spectrum obtained here exhibits the following features: i) the real part of the frequencies, $Re(\omega_n)$, is positive while the imaginary part, $Im(\omega_n)$, is negative, and therefore all modes are found to be stable, ii) the real part increases with $q,l,m$ and decreases with $n$, iii) the absolute value of the imaginary part increases with $n$, decreases with $l,m$, while $Im(\omega_n)$ itself as a function of $q$ acquires a minimum value around $q_{1,*} \simeq 0.6$.

%%%%%%%%%%%%%%%%%%%%%%%%%%%%%%%%%%%%%%%%%%%%%%%%%%%%%%%%%%%%%%

\begin{table*}[t]
\caption{QN frequencies of EpM black holes for 
$D=5, k=5/4, l=4, n=0$ and two different masses 
of the test scalar field.}
{
\begin{tabular}{l | l | l}
		 $q$ & $m=0.001$ & $m=0.1$  \\
		\hline
		\hline
0.1  &  1.79517-0.25215 i  &  1.79650-0.25187 i   \\
0.2  &  1.81978-0.25322 i  &  1.82109-0.25294 i  \\
0.3  &  1.84633-0.25418 i  &  1.84760-0.25391 i   \\
0.4  &  1.87512-0.25498 i  &  1.87637-0.25472 i   \\
0.5  &  1.90660-0.25554 i  &  1.90781-0.25529 i   \\
0.6  &  1.94131-0.25574 i  &  1.94249-0.25550 i   \\
0.7  &  1.98003-0.25537 i  &  1.98117-0.25515 i   \\
0.8  &  2.02383-0.25409 i  &  2.02493-0.25388 i   \\
0.9  &  2.07435-0.25120 i  &  2.07540-0.25101 i   \\
1.0  &  2.13417-0.24524 i  &  2.13516-0.24507 i     
\end{tabular}
	\label{table:FirstSet}
}
\end{table*}

%%%%%%%%%%%%%%%%%%%%%%%%%%%%%%%%%%%%%%%%%%%%%%%%%%%%%%%%%%%%%%

%%%%%%%%%%%%%%%%%%%%%%%%%%%%%%%%%%%%%%%%%%%%%%%%%%%%%%%
\subsection{Quasinormal spectrum in the eikonal limit}
%%%%%%%%%%%%%%%%%%%%%%%%%%%%%%%%%%%%%%%%%%%%%%%%%%%%%%%

In the eikonal regime, $l \gg 1$, the WKB approximation becomes infinitely
accurate, and therefore one can compute the QN spectrum within the first-order 
WKB semi-analytical approach. In this limit the angular momentum term is the 
dominant one in the effective potential barrier irrespectively of the dimensionality 
of space-time
\begin{equation}
V(r) \approx \frac{f(r) l^2}{r^2} \equiv l^2 g(r)
\end{equation}
where we introduce a new function $g(r) \equiv f(r)/r^2$. It is easy to verify that the maximum of the potential is located at $r_1$ that is computed solving the following algebraic equation
\begin{equation}
2 f(r_1) - r_1 f'(r)|_{r_1} = 0
\end{equation}
Then the spectrum may be computed using the first-order formula \cite{wkb1,wkb2}
\begin{equation}
\frac{i Q(r_1)}{\sqrt{2Q''(r_1)}} = n + \frac{1}{2}
\end{equation}
where the new function $Q(r)$ is defined to be $Q(r) \equiv \omega^2 - V(r)$. If the real part of the frequencies is much larger than the imaginary part, $Re(\omega_n) \gg |Im(\omega_n)|$, then it 
is not difficult to obtain the following expression for the QNMs in the eikonal limit
\begin{equation}
\omega_{l \gg 1} = \Omega_c \: l - |\lambda_L| \left( n + \frac{1}{2} \right) i
\end{equation}
where the Lyapunov exponent, $\lambda_L$, as well
as the frequency of the null geodesics, $\Omega_c$, are computed to be
\begin{eqnarray}
\Omega_c & = & \frac{\sqrt{f(r_1)}}{r_1} \\
\lambda_L  & = & r_1^2 \sqrt{\frac{g''(r_1) g(r_1)}{2}}
\end{eqnarray}
It is worth noticing that the real part of the modes is proportional to the 
angular degree, while the imaginary part depends on the overtone only. 

\begin{figure}[ht!]
\centering
\includegraphics[scale=0.9]{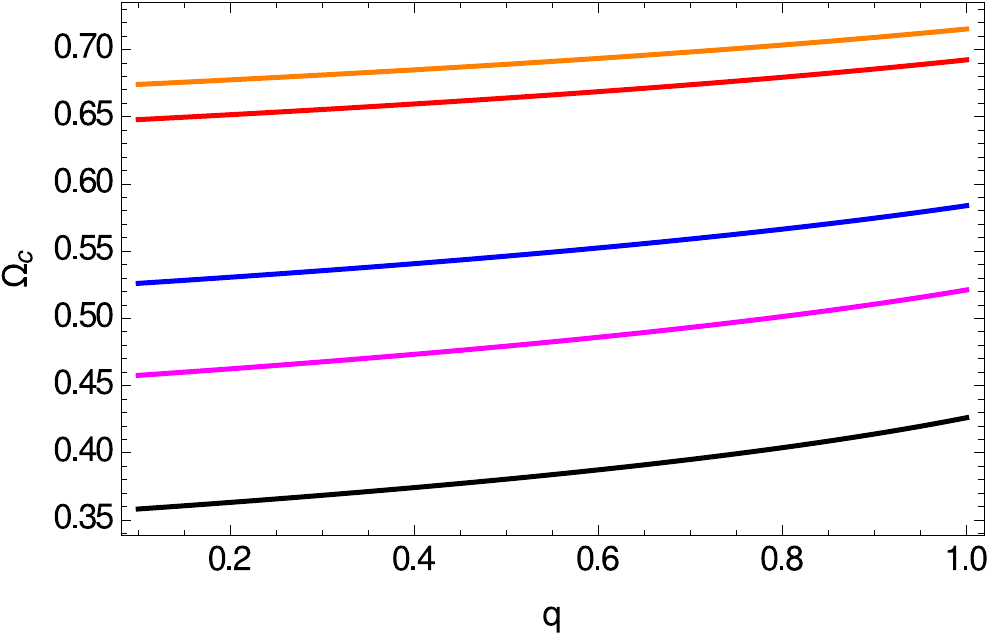} \
\includegraphics[scale=0.9]{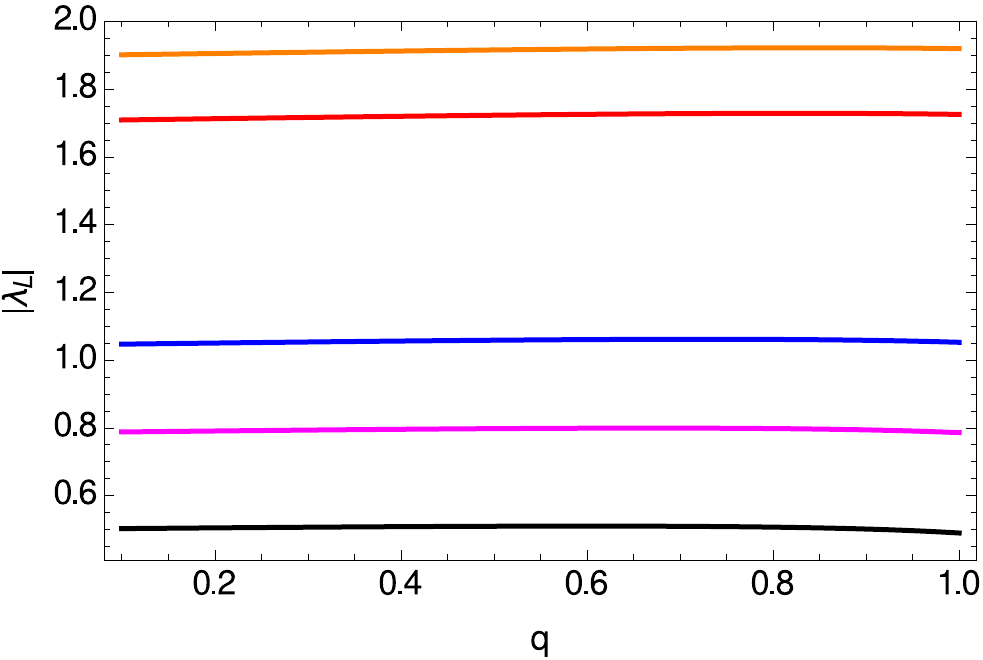}
\caption{
Eikonal regime $l \gg 1$: Frequency of null geodesics $\Omega_c$ (top panel) 
and Lyapunov exponent $|\lambda_L|$ (bottom panel) vs electric charge for $\mu=2$. 
In both panels shown are from bottom to top $D=5$ (black), $D=6$ (magenta), $D=7$ (blue), $D=10$ (red) and $D=11$ (orange).
}
\label{fig:6} 	
\end{figure}

In Fig.~\ref{fig:6} we show both $\Omega_c$ and $|\lambda_L|$ vs the electric charge
of the BH for several different dimensionalities of space-time, namely for $D=5,6,7,10$
and $D=11$, which is the maximum number of dimensions according to M-theory. 
According to our results, $\Omega_c$ is an increasing function of $q$, and 
as $D$ increases the curves are shifted upwards. Moreover, $|\lambda_L|$ exhibits 
a maximum value at a certain value of $q_{2,*}$ that depends on the dimensionality of space-time (although this feature is hardly seen in the bottom panel of Fig.~\ref{fig:6}), and the curves are again shifted upwards as $D$ increases.

Regarding future work, there are still a couple of things to be done.
For instance, one may compute the quasinormal frequencies for charged 
scalar fields, and also the quasinormal spectra for other types of fields,
such as Dirac or electromagnetic and gravitational perturbations. We hope
to be able to address those interesting issues in forthcoming works.

%%%%%%%%%%%%%%%%%%%%%%%%%%%%%%%%%%%%%%%%%%%%%%%%%%%%%%%%%%%%%%

%%%%%%%%%%%%%%%%%%%%%%%%%%%%%%%%%%%%%%%%%%%%%%%%%%%%%%%%%%%%%%%%%%%%%%%%%

\section{Conclusions}

In summary, in this work we have computed the quasinormal spectrum for scalar 
perturbations of charged five-dimensional black holes in the Einstein-power-Maxwell non-linear electrodynamics. The test field that perturbs the gravitational background is taken to be a real, massive, electrically neutral canonical scalar field, and we have adopted the popular and widely used WKB semi-analytical method. All modes are found to be stable. The mass parameter of the black hole is set to two, $\mu=2$, and we have considered two distinct values of the mass of the test scalar field ($m=0.001, 0.1$). We have shown graphically both the real and the imaginary part of the frequencies vs the electric charge of the black hole for several different values of the angular degree $l$ as well as the overtone number $n$. Finally, the QN spectrum in the eikonal regime for several different space-time dimensionalities has been computed as well.

%%%%%%%%%%%%%%%%%%%%%%%%%%%%%%%%%%%%%%%%%%%%%%%%%%%%%%%%%%%%%%%%%%%%%%%%%%%%%%

\section*{Acknowlegements}

The author thanks the Funda\c c\~ao para a Ci\^encia e Tecnologia (FCT), 
Portugal, for the financial support to the Center for Astrophysics and 
Gravitation-CENTRA,  Instituto Superior T\'ecnico,  Universidade de Lisboa,  
through the Grant No. UIDB/00099/2020.

%%%%%%%%%%%%%%%%%%%%%%%%%%%%%%%%%%%%%%%%%%%%%%%%%%%%%%%%%%%%%%%%%%%%%%%%%%%%%%

\end{document}